\documentclass[12pt]{article}
\usepackage{amssymb}

\def\Z{\mathbb Z}
\def\N{\mathbb N}
\def\a{\alpha}
\def\b{\beta}
\def\d{\delta}
\def\t{\theta}
\def\e{\varepsilon}
\def\g{\gamma}
\def\l{\lambda}
\def\r{\rho}
\def\s{\sigma}
\def\L{{\cal L}}
\def\sumjeps{\sum_{\tiny \begin{array}{c}j=1\\\e=\pm 1\end{array}}^N}
\def\Sumkjnot{\sum_{\tiny \begin{array}{c}j,k=-N \\k\not= j\end{array}}^N}
\def\Sumknotj{\sum_{\tiny \begin{array}{c}k=-N,\\k\not= j\end{array}}^N}
\def\Sumknotm{\sum_{\tiny \begin{array}{c}k=-N,\\k\not= m\end{array}}^N}
\def\prodknotj{\prod_{\tiny \begin{array}{c}k=1,\\k\not= j\end{array}}^N}
\def\Prodknotj{\prod_{\tiny \begin{array}{c}k=-N,\\k\not= j\end{array}}^N}
\def\Prodknotm{\prod_{\tiny \begin{array}{c}k=-N,\\k\not= m\end{array}}^N}
\def\Prodknoti{\prod_{\tiny \begin{array}{c}k=-N,\\k\not= i\end{array}}^N}
\def\be{\begin{equation}}
\def\ee{\end{equation}}
\def\bea{\begin{eqnarray}}
\def\eea{\end{eqnarray}}
\def\pd{\partial}    
\def\pdl{\partial\,\ln}   
\def\ds{\displaystyle}
\def\Bigdot{\begin{picture}(5,5)(0,-2)\circle*{5}\end{picture}}
 
\def\entete{
\begin{flushright}
{PAR-LPTHE 01-24} \\
{LPTM-cergy}\\
{Mai 2001}
\end{flushright}
\vspace*{5mm}
}

\def\addJean{LPTHE Paris VI (CNRS-UA 280), Box 126, Univ. Paris 6, 4 Place
Jussieu, F-75252, Paris, France}

\def\monAdd{LPTM Universit\'e de Cergy-Pontoise (CNRS ESA 8089), Neuville III, 5 mail Gay Lussac, Neuville sur Oise, F-95031 Cergy-Pontoise Cedex, France}

\newcommand{\titre}[5]{
\begin{center}
{\bf #1}\\
[10mm]
by\\
[5mm]

#2\footnote{#3} 
and #4\footnote{#5} 
\vskip 1.0in

{\bf ABSTRACT}
\end{center}
\vspace*{3mm}
}

\newtheorem{theorem}{Theorem}

\begin{document}
\pagestyle{empty}
\entete

\titre{Structures in BC$_N$ Ruijsenaars-Schneider models}{J. AVAN}{\addJean}{G. ROLLET}{\monAdd}

\noindent 
We construct the classical $r$-matrix structure for the Lax formulation of 
BC$_N$ Ruijsenaars-Schneider systems proposed in~\cite{CHY}.
The $r$-matrix structure  takes a quadratic form similar to the A$_N$  
Ruijsenaars-Schneider Poisson bracket behavior,
although the dynamical dependence is more complicated.
Commuting Hamiltonians stemming from the BC$_N$ Ruijsenaars-Schneider Lax 
matrix are shown to be linear combinations of particular 
Koornwinder-van Diejen ``external fields'' Ruijsenaars-Schneider 
models, for specific values of the exponential one-body couplings.
Uniqueness of such commuting Hamiltonians is established once the first
of them and the general analytic structure are given.
\vfill
\newpage
\setcounter{page}{1}
\pagestyle{plain}
\renewcommand{\baselinestretch}{1.5}

\section{Introduction}
The relativistic extensions of Calogero-Moser $N$-body integrable systems, 
originally introduced by Ruijsenaars and Schneider~\cite{RS}, 
have been the subject of numerous investigations in these last years.
Their exact connection to field-theoretical integrable systems, 
initially described in~\cite{RS}, was clarified in~\cite{BB}; 
their dynamical classical $r$-matrix structure, first tackled 
in~\cite{AR,NS} was finally established in~\cite{Su} and characterized as
a quadratic structure, \`a la Sklyanin~\cite{Skly,LP}, stemming from 
the dynamical linear
$r$-matrix structure of Calogero-Moser systems~\cite{AT}.
This in turn is connected to the realization of such systems~\cite{GN} 
as ``Hamiltonian reduction'', in a more general sense see~\cite{STS}, of dynamical systems living on a Heisenberg 
double~\cite{STS,Dr}, 
where
the quadratic $r$-matrix structure is natural~\cite{STS}.
More complete descriptions may be found in~\cite{VD1}.

More recently several extensions of the Ruijsenaars-Schneider systems were 
considered.
First of all, one-body-potential (or ``external field extensions'') were 
constructed in~\cite{VD2}.
The integrability proof, and construction of the quantum Hamiltonians 
themselves, rested upon polynomial-algebraic arguments~\cite{VD3} 
pioneered by Koornwinder~\cite{Ko} and indicating a connection to BC$_N$-type 
algebras.
Both Lax formulation and classical $r$-matrix construction were lacking.
Quantum elliptic van Diejen-type Hamiltonians were then constructed 
in~\cite{Kom} using corner-transfer-matrix methods, and incidentally 
pointing again to a BC$_N$ structure underlying at least some particular 
Koornwinder-van Diejen potentials.
The corner-transfer-matrix method used in~\cite{Kom} however does not 
exhibit a clear-cut way of defining a classical limit with one single Lax 
matrix, using as it does \underline{two} types of Lax operators.
Finally an explicit construction of classical BC$_N$ and C$_N$ 
Ruijsenaars-Schneider models was presented in~\cite{CHY}, 
using a $\Z_2$-folding of the original A$_{2\,N}$ or A$_{2\,N+1}$ algebra
intrinsically present in original Ruijsenaars-Schneider models.
The Lax matrix and classical commuting Hamiltonians were then constructed, 
again as $\Z_2$-foldings of the A$_{2\,N}$ (A$_{2\,N+1}$) Lax matrix and 
commuting Hamiltonians.
The elliptic BC$_N$ and C$_N$ systems were then defined in~\cite{CHY2}, and the D$_N$ case was considered in~\cite{CH}.

This series of results still begs several questions
and we wish to address here two points which remained unclarified.

First of all we construct the classical $r$-matrix for hyperbolic BC$_N$ models
(we expect that the C$_N$ and D$_N$ case, and the elliptic potentials, may be
treated by similar techniques although the elliptic case may endow more 
complicated algebraic manipulations).
This problem may seem academic, since in any case commutation of the
Hamiltonians follows from the construction itself~\cite{CHY} .
However it actually sheds light on the delicate question of 
interplay between the folding procedure A$_{2\,N}\;\rightarrow$ BC$_N$ and the 
initial ``quadratization'' of Poisson structure entailed by the change of base
symplectic manifold for the ``Hamiltonian'' reduction from $T^{*}\,g$
(cotangent bundle Lie group) to $D\,g_H$ (Heisenberg double of Lie 
group)~\footnote [1] {Similar difficulties occur when considering the quantum
deformation of BC- or D-type algebras, compared with deformation of A$_N$
algebras.
This fact was pointed out to us by D. Arnaudon.}.
Indeed it will eventually turn out that the classical $r$-matrix for 
hyperbolic  Ruijsenaars-Schneider BC$_N$ models may be recast under a 
quadratic form, similar to the A$_N$ case, but contrary to what occurred in
the Calogero-Moser models~\cite{ABT} the structure (after folding) exhibits 
now a dependance in both sets of conjugated dynamical variables, and the 
quadratic $r$-matrix is thus not directly related to the linear dynamical
$r$-matrix structure for BC$_N$ Calogero-Moser models. 

The second problem which we consider here deals with the connection between
the initial Koornwinder-van Diejen Hamiltonians and the ``canonical''
Poisson commuting Hamiltonians generated by the traces of powers of the Lax 
matrix for BC$_N$ systems.
It will be shown that the Koornwinder-van Diejen Hamiltonians are in fact 
combinations of the ``canonical ones'', and this property is actually true for 
any set of Poisson-commuting Hamiltonians with the same functional structure 
(to be explicited hereafter).

\section{The classical $r$-matrix structure}\label{rmat}
\subsection{The BC$_N$ Ruijsenaars-Schneider models and notations}
The canonical variables are a set of rapidities $\{ \t_i, i=1\cdots N\}$
and conjugate positions $q_i$ such that $\{ \t_i, q_j\} = \delta_{ij}$.
The Hamiltonian reads:
\bea\label{ham}
H & = & \sumjeps  e^ {- \e \,\b \, \t_j} f_j  + \;{\cal U}
\\
\mbox {where}\; f_j & = &  \left[ f(q_j)\,f(2\,q_j)\,\prodknotj \,f(q_j - q_k)\,f(q_j + q_k)\right]^{1/2} \quad
\mbox {and}\quad
{\cal U}  =  \prod_{k=1}^N\,f(q_k).\nonumber
\eea
Function $f$ may take different forms, namely:
\bea
f(q) & = & 1 - {g^2\over q^2} \,\, ({\rm rational})\nonumber
\\
f(q) & = & 1 - {{\rm sinh}^2 \g \over {\rm sinh}^2 {\nu q\over2}}\,\,{\rm (hyperbolic)}\nonumber
\\
f(q) & = & 1 - {{\rm sin}^2 \g \over {\rm sin}^2 {\nu q\over2}}\,\,{\rm (trigonometric)}.\nonumber
\eea
The most general elliptic case where:
\[
f (q)  =  \left( \lambda + \nu {\cal P} (q) \right) \,\, ,
\,\,{\cal P} = {\rm Weierstrass\,\, function}
\]
will not yet be considered here.

Since the trigonometric and hyperbolic cases define the same model at least
locally up to a redefinition of the parameters~\footnote [2]{
The global structure of trigonometric vs hyperbolic models is however quite different,owing to topological properties, as can be seen for instance 
in~\cite{Go}}, 
and the rational case is obtained by an easy limit procedure
from one of these models, we shall consider in the following only the hyperbolic version.

Let us note that one can also write $f(q)=v(q)\,v(-q)$, with :
\[v(q)={{\rm sinh}({{\nu q}\over{2}} +\g) \over {\rm sinh}{\nu q\over2}}\]
or even as a rational fonction of an exponential variable:
\[v(q)=\l^{-1/2}{{z-\l}\over{z-1}}\quad \mbox {with}\quad z=e^{\nu\,q}\quad \mbox {and}\quad \l=e^{-2\,\g}.\]
This rational formulation will be useful to establish some functional identities \`a la Liouville~\cite{GN}.

\subsection{The BC$_N$ Lax operator}
As shown in~\cite{CHY}, the Lax formulation of BC$_N$ Ruijsenaars-Schneider 
system may be obtained as a folding of the A$_{2n}$ case.
The reduction works as follow:
labelling the $2\,N+1$ rapidities $\{ \t_i, i=-N\cdots N\}$ and conjugate positions  $\{ q_i, i=-N\cdots N\}$, 
one sets $\t_i=\e_i\,\t_{|i|}$ and $q_i=\e_i\,q_{|i|}$ with:
\[\e_i=\left\{\begin{array}{l}
+1 \quad \mbox{for}\quad 1\le i \le N\\
0 \quad \mbox{for}\quad  i=0 \\
-1 \quad \mbox{for}\quad -1\ge i \ge -N
\end{array}\right..\]
The Lax matrix for the A$_{2n}$ cases is:
\bea
\L &=&\sum_{i,j=-N}^N \, \L_{ij} \, e_{ij} \nonumber\\\nonumber\\
\L_{ij}(q_1,...,q_N,q_0,q_{-1},...,q_{-N},\t_j)
&=&c(q_i-q_j)\,e^{-\b\,\t_j}\,\Prodknotj\,f^{1/2}(q_j-q_k)\label{LaxA2n}
\eea
where $\{ e_{ij} \}$ is the standard basis for $(2\,N+1) \times (2\,N+1)$ matrices; $f$ was
given in the previous subsection and
\[c(q)={{\rm sinh}\g \over {\rm sinh}({{\nu q}\over{2}} +\g)}
=(1-\l)\,{{z^{1/2}}\over{z-\l}}.\]
The Lax matrix for the BC$_N$ Ruijsenaars-Schneider systems then reads:
\be
L =\sum_{i,j=-N}^N \, L_{ij} \, e_{ij} \quad \mbox{with} \quad
L_{ij}=\L_{ij}(q_1,...,q_N,0,-q_1,...,-q_N,\e_j\,\t_{|j|})\label{LaxBCn}
\ee
It can be rewritten:
$\displaystyle
L_{ij}=c(q_i-q_j)\,e^{-\b\,\e_j\,\t_{|j|}}\,f_j \nonumber
$,
extending the definition of $f_j$ given in~(\ref{ham}) to $j \in \{-N...N\}$.

\noindent
Note that with this extension of $f_j$ one has $f_0={\cal U}$ and $f_j=f_{-j}$.

It has been shown that the Lax operator~(\ref{LaxA2n}) satisfies the
quadratic fundamental Poisson bracket~\cite{Su}:
\be
\left\{ \L \stackrel{\otimes}{,}\L\right\}=
\L\otimes\L\;a_{1}-a_{2}\;\L\otimes\L+\L_{2}\,s_{1}\,\L_{1}-\L_{1}\,s_{2}\,\L_{2}, 
\label{quadpoiss}
\ee
where \ $\L_{1}=\L\otimes \emph{1}$, $\L_{2}=\emph{1}\otimes \L $ and the 
quadratic structure coefficients read:
\bea
a_{1} &=&a+w,\quad s_{1}=s-w,  \nonumber \\
a_{2} &=&a+s-s^{\pi }-w,\quad s_{2}=s^{\pi }+w.\nonumber 
\eea
For any matrix $M$, the matrix $M^{\pi}$ is defined by:
\[
\mbox{if}\; M \equiv \sum_{ijkl=-N}^N \, M_{ijkl} \, e_{ ij} \otimes e_{kl} 
\quad \mbox{then}\;
M^{\pi} = \sum_{ijkl=-N}^N \, M_{ijkl} \, e_{kl}\otimes e_{ij}.
\]
Matrices $a,s,w$ take the form:
\bea\label{quadAn}
a &=&\a\,\Sumkjnot\,\mathrm{coth}{{\nu}\over{2}}(q_{k}-q_{j})\, e_{jk}\otimes e_{kj},  \nonumber
\\
s &=&-\a\,\Sumkjnot\,\frac{1}{\mathrm{sinh}{{\nu}\over{2}}(q_{k}-q_{j})}\, e_{jk}\otimes e_{kk},
\nonumber \\
w &=&\a\,\Sumkjnot\,\mathrm{coth}{{\nu}\over{2}}(q_{k}-q_{j})\, e_{kk}\otimes e_{jj}
\eea
where $\displaystyle \a \equiv \b\,{{\nu}\over{2}}$.

It must be recalled here that the most general structure of Poisson bracket
for a Lax operator of a Liouville-integrable system is a linear one~\cite{BV}:
\be\label{linpoiss}
\left\{ L \stackrel{\otimes}{,}L\right\}=
[r,L_1]-[r^{\pi},L_2].
\ee
The quadratic form~(\ref{quadpoiss}) corresponds to the case where the 
$r$-matrix itself assumes a linear dependency in $L$ of type:
\be\label{linr}
r=b\,L_2+L_2\,c,
\ee
with $b$ and $c$ arbitrary matrices determining the quadratic coefficients 
$a_1, a_2, s_1, s_2$.:
\[
a_1=c^{\pi}-c,\;a_2=b^{\pi}-b,\;s_1=c+b^{\pi} \; \mbox{and} \; s_2=s_1^{\pi}.
\]
In the next subsection, we will show that the BC$_N$ Ruijsenaars-Schneider Lax 
operator~(\ref{LaxBCn}) also satisfies a quadratic fundamental Poisson 
bracket~(\ref{quadpoiss}) albeit with a fundamental difference with respect 
to~(\ref{quadAn}) regarding the dependence on the dynamical
variables. We will give explicitely the generalizations of the
matrices $a_1,\,a_2,\,s_1$ and $s_2$, hereafter denoted ``quadratic $r$-matrices''
for obvious semantic reasons.
\subsection{Computation of the classical $r$-matrix}
Let us calculate the Poisson brackets of the Lax matrix~(\ref{LaxBCn}):
\[
\left\{ L_{ij} , L_{kl} \right\} = \b\,L_{ij} L_{kl} \left(
\e_l{{\pdl L_{ij}}\over{\pd q_{|l|}}}
-\e_j{{\pdl L_{kl}}\over{\pd q_{|j|}}}\right)
\]
and express it in terms of the  Lax matrix~(\ref{LaxA2n}),
\[
\e_l{{\pdl L_{ij}}\over{\pd q_{|l|}}}=\left(
{{\pdl \L_{ij}}\over{\pd q_l}}
-{{\pdl \L_{ij}}\over{\pd q_{-l}}}\right).
\]
We thus get:
\[
\left\{ L_{ij} , L_{kl} \right\} = 
\left\{ \L_{ij} , \L_{kl} \right\}
+\b\,L_{ij} L_{kl} \left(
{{\pdl \L_{kl}}\over{\pd q_{-j}}}
-{{\pdl \L_{ij}}\over{\pd q_{-l}}}\right).
\]
The Poisson bracket  of the first term
on r.h.s. keeps the same form~(\ref{quadpoiss}) 
where one should fold the dynamical variables
($\t_i=\e_i\,\t_{|i|}$ and $q_i=\e_i\,q_{|i|}$).
We thus only need to concentrate on the remaining term, introducing the 
four-index object:
\[U_{ijkl} \equiv {{2}\over{\nu}}\,\left(
{{\pdl \L_{kl}}\over{\pd q_{-j}}}
-{{\pdl \L_{ij}}\over{\pd q_{-l}}}\right).
\]
Straightforward calculations yield:
\[U_{ijkl}=\d_{j,-l}\,u_j+(\d_{i,-l}-\d_{j,-l})\,t_{ij}-(\d_{j,-k}-\d_{j,-l})\,t_{kl}\]
where
\bea
t_{ij}&=&-{{2}\over{\nu}}\,(\ln c)'(q_i-q_j)
=\mathrm{coth}({{\nu}\over{2}}(q_{i}-q_{j})+\g)
={{z_i+\l\,z_j}\over{z_i-\l\,z_j}}
\nonumber\\
u_j&=&\!\!\!\!{{2}\over{\nu}}\,\Sumknotj\,(\ln f)'(q_k-q_j)
\nonumber\\
&&=\Sumknotj\,2\,\mathrm{coth}{{\nu}\over{2}}(q_{j}-q_{k})
+\mathrm{coth}({{\nu}\over{2}}(q_{k}-q_{j})+\g)
-\mathrm{coth}({{\nu}\over{2}}(q_{j}-q_{k})+\g)
\nonumber\\
&&=\Sumknotj\,2\,{{z_j+z_k}\over{z_j-z_k}}+{{z_k+\l\,z_j}\over{z_k-\l\,z_j}}-{{z_j+\l\,z_k}\over{z_j-\l\,z_k}}
\,=\,\Sumknotj\,2\,a_{jk}+t_{kj}-t_{jk},
\nonumber
\eea
with $\ds a_{jk}=\mathrm{coth}{{\nu}\over{2}}(q_{j}-q_{k})=
{{z_j+z_k}\over{z_j-z_k}}$ for $j \ne k$.

Note the following properties of these objects on the folded space:
\[
t_{-j-i}=t_{ij},\,a_{-j-i}=a_{ij},\,\mbox{ and } \; u_{-i}=-u_i.
\]
In the expression of $U_{ijkl}$, the terms in $\d_{j,-l}$ 
are not on the same footing as the others since they are separately 
antisymmetric under the exchange of the two spaces (operation ${\pi}$) 
whereas the remaining terms verify this property only altogether.

We thus first take care of the $\d_{j,-l}$ terms, introducing the matrix $\r$:
\[
\r=\a\,\sum_{k,l=-N }^N\,L_{kl}\,({{1}\over{2}}\,u_l-t_{kl})\,e_{-l-l}\otimes e_{kl},
\]
realizing them as a linear $r$-matrix form~(\ref{linpoiss}):
\bea
\left([\r ,L_{1}] - [\r^{\pi} ,L_{2}]\right)_{ijkl}=
\a\,L_{ij} L_{kl}\,\left (\d_{j,-l}\,(u_j -t_{ij}+t_{kl})
+\d_{i,-l}\,({{1}\over{2}}\,u_l-t_{kl})-\d_{j,-k}\,({{1}\over{2}}\,u_j-t_{ij})
\right )\nonumber\\
=\a\,L_{ij} L_{kl}\,(U_{ijkl}-\tilde U_{ijkl}),
\quad \mbox{with} \;\; \tilde U_{ijkl}=
\d_{i,-l}\,(t_{ij}+t_{kl}-{{1}\over{2}}\,u_l)-\d_{j,-k}\,(t_{ij}+t_{kl}-{{1}\over{2}}\,u_j).
\nonumber
\eea
We may furthermore bring it back to our seeked-for general quadratic form by setting
$\r=\tau\,L_2$ (i.e. taking $b=\tau$ and $c=0$ in~(\ref{linr})), since the matrix $L$ is
invertible: 
\be\label{tau}
\tau=\r\,L_2^{-1}=
\a\,\sum_{i,k,l=-N }^N\,L_{k-i}\,L_{-il}^{-1}\,({{1}\over{2}}\,u_{-i}-t_{k-i})\,e_{ii}\otimes e_{kl}.
\ee
One should immediately note, from the explicit form of $L$~(\ref{LaxBCn}),
that this matrix $\tau$ actually does not depend on the rapidities
$\t_i$'s.
We are therefore still in the ``canonical'' quadratic structure (a la Suris) of 
type~(\ref{quadpoiss})
with dynamical quadratic $r$-matrices depending only on one set of canonical 
variables (the $q_i$'s).

We will now show that  there exists a matrix $\s$,
such that $\s^{\pi}=\s$ and:
\[
\a\,L_{ij} L_{kl}\, \tilde U_{ijkl}=\left[
L_{2}\,\s\, L_{1}- L_{1}\,\s^{\pi}\,L_{2}\right]_{ijkl}.
\]
This corresponds to setting $c=\s$ and $b=0$ in~(\ref{linr}) and thus actually
formally completes the quadratic $r$-matrix structure~(\ref{quadpoiss}).

In order to ensure self-consistency of the dependence in
the indices,
we assume the following tensorial structure:
\[
\s=\sum_{m,n=-N }^N\,\s_{mn}\,e_{mn}\otimes e_{-n-m},
\quad \mbox{satisfying} \quad \s^{\pi}=\s \quad \mbox{ie:}
\quad \s_{-n-m}=\s_{mn},
\]
yielding:
\[
\left[L_{2}\,\s\, L_{1}- L_{1}\,\s^{\pi}\,L_{2}\right]_{ijkl}
=\d_{i,-l}\,\sum_{n=-N }^N\,\s_{in}\,L_{k-n}\,L_{nj}
-\d_{j,-k}\,\sum_{n=-N }^N\,\s_{kn}\,L_{i-n}\,L_{nl}.
\]
The set of equations to be solved then reads:
\[
\d_{i,-l}\,\left(
\sum_{n=-N }^N\,\s_{in}\,{{L_{k-n}\,L_{nj}}\over{L_{ij} L_{kl}}}
-\a\,(t_{ij}+t_{kl}-{{1}\over{2}}\,u_l)\right)
=\d_{j,-k}\,\left(
\sum_{n=-N }^N\,\s_{kn}\,{{L_{i-n}\,L_{nl}}\over{L_{ij} L_{kl}}}
-\a\,(t_{ij}+t_{kl}-{{1}\over{2}}\,u_j)\right),
\]
or, equivalently:
\[
\sum_{n=-N }^N\,\s_{in}\,{{L_{k-n}\,L_{nj}}\over{L_{ij} L_{k-i}}}
-\a\,(t_{ij}+t_{k-i}+{{1}\over{2}}\,u_i)
=\d_{j,-k}\,s_{ik}, \quad \mbox{with} \quad s_{ik}=s_{ki}.
\]
Direct calculations yield:
\[
{{L_{k-n}\,L_{nj}}\over{L_{ij} L_{k-i}}}=
{{e^{\b\,\t_n}\,f_n}\over{e^{\b\,\t_i}\,f_i}}\,\left(
(1-\d_{j,-k})\,{{t_{nj}-t_{n-k}}\over{t_{ij}-t_{i-k}}}
+\d_{j,-k}\,{{t_{nj}^2-1}\over{t_{ij}^2-1}}
\right),
\]
and equations become:
\bea
\sum_{n=-N }^N\,\tilde \s_{in}\,(t_{nj}-t_{n-k}) &=&
\a\,(t_{ij}^2-t_{i-k}^2+{{1}\over{2}}\,u_i\,t_{ij}
-{{1}\over{2}}\,u_i\,t_{i-k})
\nonumber\\
\sum_{n=-N }^N\,\tilde \s_{in}\,{{t_{nj}^2-1}\over{t_{ij}^2-1}}
-\a\,(2\,t_{ij}+{{1}\over{2}}\,u_i) &=& s_{i-j},
\quad \mbox{with} \quad
\tilde \s_{in}={{e^{\b\,\t_n}\,f_n}\over{e^{\b\,\t_i}\,f_i}}\,\s_{in}.
\nonumber
\eea
Let us recall here that the only additional contraints on matrices
$\tilde \s$ and $s$ read:
\be\label{compa1}
\left \{\begin{array}{l}
\tilde \s_{-j-i}={{f_i^2}\over{f_j^2}}\,\tilde \s_{ij}
\\
s_{ij}=s_{ji}
\end{array} \right .
\ee
According to the previous equations, the matrix $s$ is determinated given the matrix
$\tilde \s$,
and $\tilde \s$ is obtained, up to a one-dimensional degree of
freedom $v_i$, by:
\be\label{sigma}
\sum_{n=-N }^N\,\tilde \s_{in}\,t_{nj}=
\a\,(t_{ij}^2+{{1}\over{2}}\,u_i\,t_{ij}+v_i),
\ee
since $t$ is invertible.

Remains to verify that one can find $v_i$'s, such that
the compatibility relations~(\ref{compa1}) be satisfied, namely:
\bea
\sum_{m=-N }^N\,t_{im}\,f_m^2\,t_{mj}(t_{mj}-t_{im}+u_m)=
\sum_{m=-N }^N\,f_m^2\,(v_{-m}\,t_{mj}-v_m\,t_{im})\label{compaa}\\
(v_i+1)\,\sum_{n,m=-N }^N\,(t_{-jm}^2-1)\,(t^{-1})_{mn}=
(v_{-j}+1)\,\sum_{n,m=-N }^N\,(t_{im}^2-1)\,(t^{-1})_{mn}\label{compab}
\eea
Equation~(\ref{compab}) directly yields
$\ds
v_i=-1+\eta\,\sum_{n,m=-N }^N\,(t_{im}^2-1)\,(t^{-1})_{mn},
$ with $\eta$ an arbitrary constant.

In order to solve~(\ref{compaa})  we shall first compute its left hand side.

We do so, by twofold evaluation of
the following contour integral in the complex plane around
infinity:
\[
I_{ij} = \frac{1}{2\pi i} \oint_{C_{\infty}} {{dz}\over{z}} 
\;{{z_i+\l\,z}\over{z_i-\l\,z}}
\;{{z+\l\,z_j}\over{z-\l\,z_j}}
\,\prod_{k=-N}^N
\;{{z-\l\,z_k}\over{z-z_k}}
\;{{z_k-\l\,z}\over{\l\,(z_k-z)}}.
\]
Contour $C_{\infty}$ is oriented counterclockwise and loops around infinity.

\noindent
Residue at infinity gives $I_{ij} =-1$.

Whereas $I_{ij}$ also equals the sum of residues at poles of the meromorph 
integrand in the whole complex plane, that is: a single pole at $z=0$ with 
residue $-1$ and a set of double poles at $z= z_m$.
We thus obtain:
\[
\sum_{m=-N }^N\,\left[
{{z_i+\l\,z}\over{z_i-\l\,z}}
\;{{z+\l\,z_j}\over{z-\l\,z_j}}
\;{{(z-\l\,z_m)\,(z_m-\l\,z)}\over{-\l\,z}}
\Prodknotm
\;{{z-\l\,z_k}\over{z-z_k}}
\;{{z_k-\l\,z}\over{\l\,(z_k-z)}}
\right]'(z=z_m)=0.
\]
Noticing that:
\bea
u_m&=&\!\!\!\!{{2}\over{\nu}}\Sumknotm\,(\ln f)'(q_k-q_m)
\,=\,-{{2}\over{\nu}} {{\pd}\over{\pd q_m}}\, \ln\,\Prodknotm \,f (q_k-q_m)
\nonumber\\
&=&- 2\,z_m\,{{\pd}\over{\pd z_m}}\, \ln\,\Prodknotm 
\;{{z_m-\l\,z_k}\over{z_m-z_k}}
\;{{z_k-\l\,z_m}\over{\l\,(z_k-z_m)}}
\nonumber\\
&=&- 2\,z_m\,\left[\ln\,\Prodknotm 
\;{{z-\l\,z_k}\over{z-z_k}}
\;{{z_k-\l\,z}\over{\l\,(z_k-z)}}
\right]'(z=z_m)
\nonumber
\eea
and also:
\[
z_m\,\left[\ln\,({{z_i+\l\,z}\over{z_i-\l\,z}}\;{{z+\l\,z_j}\over{z-\l\,z_j}})
\right]'(z=z_m)=
{{1}\over{2}}\,((t_{im}-t_{mj})-({{1}\over{t_{im}}}-{{1}\over{t_{mj}}})),
\]
we obtain:
\[
\sum_{m=-N }^N\,t_{im}\,f_m^2\,t_{mj}\left[t_{mj}-t_{im}+u_m +
({{1}\over{t_{im}}}-{{1}\over{t_{mj}}})\right]=0,
\]
or equivalently:
\[
\sum_{m=-N }^N\,t_{im}\,f_m^2\,t_{mj}(t_{mj}-t_{im}+u_m)=
\sum_{m=-N }^N\,f_m^2\,(t_{im}-t_{mj}).
\]
This derivation of Liouville-type functional identities stems from similar 
derivations to be found in~\cite{GN}.

Equation~(\ref{compaa}) now reads:
\[
\sum_{m=-N }^N\,t_{im}\,f_m^2\,(v_m+1)=\sum_{m=-N }^N\,(v_{-m}+1)\,f_m^2\,t_{mj}
=\sum_{m=-N }^N\,t_{-jm}\,f_m^2\,(v_m+1),
\]
and can be solved straightforwardly:
$\ds v_i=-1+\xi\,{{1}\over{f_i^2}}\,\sum_{m=-N }^N\,(t^{-1})_{im}$,
with $\xi$ any constant.

We thereby prove the consistency of~(\ref{compa1}) and~(\ref{sigma}), since $v_i=-1$
is an obvious solution (with $\eta=\xi=0$).

In addition, performing calculations of the same type of contour integrals, one gets:
\[
\sum_{n,m=-N }^N\,(t_{im}^2-1)\,(t^{-1})_{mn}=
-{{4\,\l}\over{(\l-1)^2}}\,{{1}\over{f_i^2}}\,\sum_{m=-N }^N\,(t^{-1})_{im}
\propto {{1}\over{D^+_i}},
\]
with $\ds D_i^+ =\,\Prodknoti\,{{z_i-\l\,z_k}\over{z_i-z_k}}$.

The two forms which the $v_i$'s should satisfy, are actually identical:
$\ds v_i=-1+{{\zeta}\over{D^+_i}}$, $\zeta$ being an arbitrary
constant.

We fix this remaining gauge, setting
$\ds \zeta={{1-\l^{2\,N+1}}\over{\!\!\!\!1-\l}}$,
in order to obtain the simplest form for $\tilde \s$
from relation~(\ref{sigma}):
\[
\tilde \s_{ij}=
\a\,{{D_j^+}\over{D_i^+}}\,
\left (\d_{i,j}\,s_i-(1-\d_{i,j})\,a_{ij}\right ),
\quad \mbox{where} \quad 
s_i={{1+\l}\over{1-\l}}\,+\,\sum_{m=-N }^N\,{{1}\over{2}}\,(t_{mi}+t_{im}).
\]
We finally give the expression of the matrix $\sigma$:
\be\label{sig}
\s=\a\,\sum_{i,j=-N }^N\,
{{{\cal A}_j}\over{{\cal A}_i}}\,
\left (\d_{i,j}\,s_i-(1-\d_{i,j})\,a_{ij}\right )\,
\,e_{ij}\otimes e_{-j-i},
\quad \mbox{with} \quad
{\cal A}_i=\sqrt{{{D_i^+}\over{D_{-i}^+}}}\,e^{-\b\,\t_i}.
\ee

The $r$-matrix structure is now completely defined by a quadratic Poisson bracket of 
type~(\ref{quadpoiss}) where the quadratic $r$-matrices $a_1,\,a_2,\,s_1$ and $s_2$ 
are changed into:
\bea\label{newquadpoiss}
a_1\,&\rightarrow & \tilde{a}_1=a_1, \nonumber\\
s_1\,&\rightarrow & \tilde{s}_1=s_1+\sigma+\tau^{\pi}\nonumber\\
s_2\,&\rightarrow & \tilde{s}_2=\tilde{s}_1^{\pi}
=s_2+\sigma^{\pi}+\tau=s_2+\sigma+\tau\nonumber\\
a_2\,&\rightarrow & \tilde{a}_2=\tilde{a}_1+\tilde{s}_1-\tilde{s}_2
=a_2+\tau^{\pi}-\tau,
\eea
and matrices $\sigma$ and $\tau$ are respectively defined by
equations~(\ref{sig}) and ~(\ref{tau}).

\subsection{Comments}
One should notice that this quadratic $r$- matrix structure is now
fully dynamical, depending both
on the positions $q_i$'s and rapidities $\t_i$'s.
Moreover, its conjugating factor ${\cal A}_i$, which bears this double dependance,
is deeply connected to the structure of the matrix $L$ under folding:
\[
L^{-1}_{ij}={{{\cal A}_j}\over{{\cal A}_i}}\,L_{-i-j}
\]
We have here an interesting first example of a ``doubly dynamical'' $r$-matrix
dependence, stemming from the interplay between the folding procedure leading
from A$_N$ to BC$_N$, and the quadratic structure of Ruijsenaars-Schneider-type
models. This seems to open new perspectives, first of all on the classification of
such doubly dynamical $r$-matrices. The only other example known to us at this time
is the classical linear $r$-matrix structure for the Lax formulation of the
$A_N$ elliptic Calogero-Moser model in the absence of spectral parameter
~\cite{OP}.
Proof of its double dynamical dependance is given in~\cite{BS} although the
explicit form is yet conjectural~\cite{Br}. Curiously however, it occurs in
relation with an $A_N$ model with no folding and may therefore be of a different 
nature.

Interpretation of  doubly dynamical objects in the frame of quantum
group theory is yet lacking. ``Simply'' dynamical $r$-matrices 
are known to be connected to the theory of Drinfel'd-twisted quantum groups, 
specifically of the type of Felder's Dynamical Quantum Groups~\cite{Fe} 
(see for instance~\cite{Ha,BBB}). Whether doubly dynamical
objects have such connections is a new problem and we have no further
comments to make on this point. A very recent result~\cite{PX} may 
however give indications on how to twist quantum groups by non-abelian 
twisted cocycles (here the twist would occur ``along'' a Heisenberg algebra).

\section{The canonical Hamiltonians}
\subsection{Preliminaries}

We first describe the Poisson-commuting Hamiltonians generated by
traces of powers of the BC$_N$ Lax matrix~(\ref{LaxBCn}).
They read for any integer $l\in \{1..N\}$:
\[
{\cal H}_l=tr(L^l)=\sum_{\tiny \begin{array}{c}
{\cal J}\subset \{-N..N\}\\
|{\cal J}|=l
\end{array}} m_{{\cal J}}(L),
\]
with $m_{{\cal J}}(L)$ the principal minor of $L$ with lines and columns indexed by
${\cal J}$.

Taking into account the form of $L$ and properties of Cauchy matrices:
\[
m_{{\cal J}}(L)=e^{-\b \t_{{\cal J}}}
\prod_{\tiny \begin{array}{c} 
j\in {\cal J}\\ k\notin {\cal J}
\end{array}} f^{1/2}(q_j-q_k),
\quad \mbox{where} \quad \t_{{\cal J}}=\sum_{j\in {\cal J}} \t_j.
\]
Because of the folding, we now rearrange these terms so as to sort them 
with respect to linearly independent  exponentials of rapidities.
We thus decompose  ${\cal J}=\e J \bigcup  S$, separating indices of ${\cal J}$
such that their opposite does not belong to ${\cal J}$
(set $\e J=\{\e_j |j|/j \in {\cal J} \wedge -j \notin {\cal J}\}$ and 
$J=\{|j|/j \in \e J \}\subset \{1..N\}$) 
and the complementary parts, symmetric under folding $S$:
\be\label{calHl}
{\cal H}_l=\hskip -.5cm
\sum_{\tiny \begin{array}{c}
J\subset \{1..N\},\,|J|\le l\\
\e_j=\pm 1, \, j \in J
\end{array}} {\cal U}_{J^c,l-|J|}\quad e^{-\b \t_{\e J}}
\prod_{\tiny \begin{array}{c} 
j\in \e J\\ k\notin \e J
\end{array}} f^{1/2}(q_j-q_k),
\ee
with
\be\label{calU}
{\cal U}_{K,p}=\hskip -.5cm
\sum_{\tiny \begin{array}{c}
S\subset {\cal A}_{K}=K \bigcup -K \bigcup \{0\}\\
S=-S,\,|S|=p
\end{array}}\hskip -.25cm
\prod_{\tiny \begin{array}{c} 
s\in S\\ k\in {\cal A}_{K}\backslash S
\end{array}}\hskip -.25cm f^{1/2}(q_s-q_k)
=\hskip -.5cm
\sum_{\tiny \begin{array}{c}
S\subset {\cal A}_{K}\\
S=-S,\,|S|=p
\end{array}}\hskip -.25cm
\prod_{\tiny \begin{array}{c} 
s\in S\\ k\in {\cal A}_{K}\backslash S
\end{array}}\hskip -.25cm v(q_s-q_k).
\ee
We then recall the  Koornwinder-van Diejen Hamiltonians~\cite{VD2} in the classical case:
\be\label{Hl}
H_l=
\sum_{\tiny \begin{array}{c}
J\subset \{1..N\},\,|J|\le l\\
\e_j=\pm 1, \, j \in J
\end{array}} \hskip -.25cm U_{J^c,l-|J|}\quad e^{-\b \t_{\e J}}\quad
V^{1/2}_{\e J; J^c}\quad V^{1/2}_{-\e J; J^c},
\ee
where, after some rearrangements,
\[
V_{\e J; K}=\prod_{j \in \e J} {{w(q_j)}\over{v(2\,q_j)\,v(q_j)}}
\prod_{\tiny \begin{array}{c}
j \in \e J \\ k \in {\cal A}_K \bigcup -\e J
\end{array}} \hskip -.5cm v(q_j-q_k)
\]
and
\[
U_{K,p}=(-1)^p 
\sum_{\tiny \begin{array}{c}
\e I \subset {\cal A}_{K}\\
|I|=p
\end{array}}
\prod_{i \in \e I} {{w(q_i)}\over{v(2\,q_i)\,v(q_i)}}
\prod_{\tiny \begin{array}{c}
i, i' \in \e I\\i < i'
\end{array}} \hskip -.25cm
{{v(-q_i-q_{i'})}\over{v(q_i+q_{i'})}}
\prod_{\tiny \begin{array}{c} 
i\in \e I\\ k\in {\cal A}_{K}\backslash \e I
\end{array}} \hskip -.5cm v(q_i-q_k).
\]
The $w$ are particular functions and may be interpreted as an interaction 
with some external field.

Direct computation yields:
\[
V_{\e J; J^c}\; V_{-\e J; J^c}=
\prod_{j \in \e J} {{w(q_j)}\over{v(2\,q_j)\,v(q_j)}}\,
{{w(-q_j)}\over{v(-2\,q_j)\,v(-q_j)}}
\prod_{\tiny \begin{array}{c} 
j\in \e J\\ k\notin \e J
\end{array}} f(q_j-q_k).
\]
Setting $w(q_j)=v(2\,q_j)\,v(q_j)$, which is an admissible choice according 
to~\cite{VD2}, $H_l$~(\ref{Hl}) takes actually the same form as 
${\cal H}_l$~(\ref{calHl}), up to
the change of ${\cal U}_{K,p}$ in to $U_{K,p}$.
$U_{K,p}$ takes a simpler form, for this one body potential $w$:
\be\label{U}
U_{K,p}=(-1)^p 
\sum_{\tiny \begin{array}{c}
\e I \subset {\cal A}_{K}\\
|I|=p
\end{array}}
\prod_{\tiny \begin{array}{c}
i, i' \in \e I\\i < i'
\end{array}} \hskip -.25cm
{{v(-q_i-q_{i'})}\over{v(q_i+q_{i'})}}
\prod_{\tiny \begin{array}{c} 
i\in \e I\\ k\in {\cal A}_{K}\backslash \e I
\end{array}} \hskip -.5cm v(q_i-q_k),
\ee
nevertheless {\it it is not generally equal to} ${\cal U}_{K,p}$ (the notation
used in~\cite{CHY} is in this respect misleading).

They are actually only equal for $p=0$, where trivially:
$U_{K,0}=1={\cal U}_{K,0}$.
For instance when $p=1$, one gets:
\[
U_{K,1}=-\sum_{i \in {\cal A}_{K}\backslash \{0\}}
\prod_{\tiny \begin{array}{c}  k\in {\cal A}_{K}\\ k\not=i\end{array}} \hskip -.25cm v(q_i-q_k) \quad \mbox{and} \quad 
{\cal U}_{K,1}=
\prod_{\tiny \begin{array}{c}  k\in {\cal A}_{K}\\ k\not=0\end{array}} \hskip -.25cm v(q_k).
\]
We compute a suitable contour integral on the same lines as in the previous section
to obtain the Liouville-type functional identity:
\[\sum_{i \in {\cal A}_{K}}
\prod_{\tiny \begin{array}{c}  k\in {\cal A}_{K}\\ k\not=i\end{array}} \hskip -.25cm v(q_i-q_k)={{\sinh \g(2\,|K|+1)}\over{\sinh \g}},
\]
and thus show that: $U_{K,1}={\cal U}_{K,1}-{{\sinh \g(2\,|K|+1)}\over{\sinh \g}}$.

It will now be shown that these two relations, for $p=0$ and $p=1$, between
the $U_{K,p}$'s and ${\cal U}_{K,1}$'s are actually sufficient to establish 
that the two sets of Hamiltonians define the same family of commuting 
dynamical 
flows, namely one set of Hamiltonians is a triangular linear combination
of the other set.

A more general result will in fact be proved in the following subsection.

\subsection{Uniqueness theorem}

\begin{theorem}
Let $q_i$ and $\t_i$, $i\in \N$, be a set of conjugated variables
such that $\{ \t_i, q_j\} = \delta_{ij}$.
Let $I$ and $K$ be  arbitrary finite sets of indices included in $\N$.
Assume the existence of a set of complex functions  $u_{K,p}$ depending 
upon the set of
indices $K$  and a natural integer $p$, and of another set
of complex functions  $v_{\e J,I}$ depending upon the sets of
indices $J$ and $I$ ($J\subset I$) and a $|J|$-uple of signs  $\e=(\e_j,\,j\in J)$,
such that:

\Bigdot $u_{K,p}$ and $v_{\e J,I}$ be independent of the rapidities $\t_i$s.

\Bigdot $u_{K,0}=1$, $v_{\emptyset,I}=1$, and $v_{\e \{j\},I}\not\equiv 0$.

\Bigdot $\ds S^I= \{\; h_l^I=
\hskip -.5cm \sum_{\tiny \begin{array}{c}
J\subset I,\,|J|\le l\\
\e_j=\pm 1, \, j \in J
\end{array}} \hskip -.25cm u_{J^c,l-|J|}\quad e^{-\b \t_{\e J}}\quad
v_{\e J,I},\,l\in \{1..|I|\}\; \}$ be a family of Poisson-commuting functions
($\ds \t_{\e J}=\sum_{j\in J}\e_j \t_j$).

If there exists a second set of  complex functions  $\tilde u_{K,p}$
obeying the first two conditions; 
such that $\ds \tilde S^I= \{\; \tilde h_l^I=
\hskip -.5cm \sum_{\tiny \begin{array}{c}
J\subset I,\,|J|\le l\\
\e_j=\pm 1, \, j \in J
\end{array}} \hskip -.25cm \tilde u_{J^c,l-|J|}\quad e^{-\b \t_{\e J}}\quad
v_{\e J,I},\,l\in \{1..|I|\}\; \}$ be a new family of Poisson-commuting functions;
and $\tilde u_{K,1}=u_{K,1}+c_1(|K|)$,
then there exist coefficients $c_r(m)$, $(r,m)\in \N^2$,
independent of all dynamical variables,
connecting the two families of Hamiltonians as:
\[
\tilde h_l^I=\sum_{s=0}^l c_{l-s}(|I|-s)\,h_s^I,
\,\mbox{ with }\, \forall m \in \N, c_0(m)=1.
\]
\end{theorem}
{\bf Proof:}

The strategy of the proof relies upon a recursive procedure on $p$, showing that:
\be\label{rh}
\tilde u_{K,p}=\sum_{r=0}^p c_{p-r}(|K|-r)\,u_{K,r},\quad  
\forall K \subset \N
\; \mbox{ finite and such that } \; |K|\ge p.
\ee
Let $l_0$ be a strictly positive integer; the recursion hypothesis hereafter
denoted r.h.,
states that (\ref{rh}) is valid for any $p \le l_0$.

The assumptions in the theorem immediately imply the validity of r.h. for 
$l_0=1$, can be directly rewritten as:
\be\label{2h1}
\tilde h_{1}^I=h_{1}^I+c_1(|I|).
\ee
Let us assume r.h. up to $l_0$ and establish it for $l_0+1$.
We have:
\[
\tilde h_{l_0+1}^I=\tilde u_{I,l_0+1}+
\hskip -.5cm \sum_{\tiny \begin{array}{c}
J\subset I,\,1\le |J|\le l_0+1\\
\e_j=\pm 1, \, j \in J
\end{array}} \hskip -.25cm \tilde u_{J^c,l_0+1-|J|}\quad e^{-\b \t_{\e J}}\quad
v_{\e J,I}.
\]
Since $l_0+1-|J|\le l_0$ in the previous summation, we apply r.h. to get:
\bea\label{hl+1}
\tilde h_{l_0+1}^I&=&\tilde u_{I,l_0+1}+
\sum_{s=1}^{l_0+1} c_{l_0+1-s}(|I|-s)
\hskip -.5cm \sum_{\tiny \begin{array}{c}
J\subset I,\,1\le |J|\le s\\
\e_j=\pm 1, \, j \in J
\end{array}} \hskip -.25cm u_{J^c,s-|J|}\quad e^{-\b \t_{\e J}}\quad
v_{\e J,I}\nonumber\\
&=&\tilde u_{I,l_0+1}-
\sum_{s=1}^{l_0+1} c_{l_0+1-s}(|I|-s)\,u_{I,s}+
\sum_{s=1}^{l_0+1} c_{l_0+1-s}(|I|-s)\,h_s^I.
\eea
We now use the Poisson-commutation property of $\tilde S^I$ as:
\[
\{\tilde h_{l_0+1}^I,\tilde h_{1}^I\}=0.
\]
Combining (\ref{2h1}), (\ref{hl+1}) and  the Poisson-commutation property of $S^I$ yields:
\bea
0&=&\{\tilde u_{I,l_0+1}-\sum_{s=1}^{l_0+1} c_{l_0+1-s}(|I|-s)\,u_{I,s},
h_{1}^I\}\nonumber\\
&=&
\sum_{j\in I, \e=\pm 1} \b\,\e\,e^{-\b\,\e\, \t_{j}}\,v_{\e \{j\},I}
{{\pd}\over{\pd q_j}} (\tilde u_{I,l_0+1}-\sum_{s=1}^{l_0+1} c_{l_0+1-s}(|I|-s)\,u_{I,s}).\nonumber
\eea
By functional independence of $\ds \sum_{\e=\pm 1} \e\,e^{-\b\,\e\, \t_{j}}\,v_{\e \{j\},I}$, it follows that the function obtained as $\ds \tilde u_{I,l_0+1}-\sum_{s=1}^{l_0+1} c_{l_0+1-s}(|I|-s)\,u_{I,s}$ {\it does not depend on any dynamical variable}.

Hence it defines the coefficient $ c_{l_0+1}(|I|)$, thereby proving the r.h.
to order $l_0+1$.

Finally, relation~(\ref{rh}) immediately implies the result of the theorem.
$\square$

\subsection{Comments}

An immediate consequence of this theorem is the existence of linear triangular
relations between the BC$_N$ Ruijsenaars-Schneider Hamiltonians
and the classical Koornwinder-van Diejen Hamiltonians when
$w(q_j)=v(2\,q_j)\,v(q_j)$.

The explicit coefficient have to be computed order by order since at this 
time no general recursion formula is available.
As an example we have worked out the first two functions:
\bea
U_{K,1}={\cal U}_{K,1}+c_1(|K|)\quad \mbox{ and } \quad
U_{K,2}={\cal U}_{K,2}+c_1(|K|-1)\,{\cal U}_{K,1}+c_2(|K|)\nonumber \\
\nonumber \\
\mbox{ with } \quad c_1(|K|)=-{{\sinh \g(2\,|K|+1)}\over{\sinh \g}},
\hskip 5cm \nonumber \\
c_2(|K|)={{1}\over{2}}\,\left (
{{\sinh \g(2\,|K|-1)}\over{\sinh \g}}\,{{\sinh \g(2\,|K|+1)}\over{\sinh \g}}
-{{\sinh 4\,\g \,|K|}\over{\sinh 2\,\g}}-2
\right ).\nonumber 
\eea
No obvious pattern appears yet.
As a consequence, an algebraic interpretation of the Koornwinder-van Diejen 
Hamiltonians in connection with the canonical Hamiltonians is still lacking.

More generally, the theorem implies that a hierarchy of Poisson-commuting
Hamiltonians with the generic form given is uniquely determinated by the 
giving of the family of $v$-functions and the first Hamiltonian, or equivalently the first ``potential term'' $u_{K,1}$.
In the Koornwinder-van Diejen case, this first Hamiltonian is given 
in~\cite{Ko}.

We wish to end this section with a conjecture on the classical 
Koornwinder-van Diejen Hamiltonians with  a general one-body potential
chosen as in~\cite{VD2} (this time dropping the restriction to $w(q_j)=v(2\,q_j)\,v(q_j)$).
They have not yet been constructed by a Lax formalism.
We expect that the suitable Lax matrix for this hierarchy may be obtained
by multiplying the BC$_N$ Lax matrix~(\ref{LaxBCn})  by a $2\,N+1$ diagonal
matrix: $L_W=L D_W$, with $D_{W_{ii}}={\cal W}(q_i)\,{\cal W}(-q_i)$.
This one-body potential function ${\cal W}$ has to be determinated by
integrability conditions.
In addition, we conjecture that, after some canonical transformation on the dynamical variables, the Hamiltonians ${\cal H}_W^l=tr(L_W^l)$ will take the same  form as the Koornwinder-van Diejen Hamiltonians~(\ref{Hl}) up to the change
of $U_{K,p}$ into some ${\cal U}_{W_{K,p}}$.
The theorem will then apply, thereby yielding the full connection between BC$_N$-type  Ruijsenaars-Schneider potentials 
and the classical Koornwinder-van Diejen Hamiltonians.

\section*{Acknowledgements}
We wish to thank Daniel Arnaudon for pointing out to us some subtleties
regarding the interplay between folding and quadratization,
and Olivier Babelon for drawing our attention to reference~\cite{PX}.

\end{document}